\documentclass[aip,jap,reprint]{revtex4-1}
\usepackage{graphicx}
\usepackage{lipsum}
\pdfoutput=1

\begin{document}

\widetext
\leftline{The following article has been accepted by Applied Physics Letters.}
\leftline{After it is published, it will be found at \url{http://apl.aip.org/}.}

\title{Robustness of n-GaAs Carrier Spin Properties to 5 MeV Proton Irradiation}%

\author{Brennan C. Pursley}%
\email{bpursley@umich.edu}
\affiliation{Applied Physics Program, University of Michigan, Ann Arbor, MI 48109}

\author{X. Song}%
\affiliation{Applied Physics Program, University of Michigan, Ann Arbor, MI 48109}

\author{R. O. Torres-Isea}%
\affiliation{Department of Physics, University of Michigan, Ann Arbor, MI 48109}

\author{E. A. Bokari}%
\affiliation{Department of Physics, Western Michigan University, Kalamazoo, MI 49008}

\author{A. Kayani}%
\affiliation{Department of Physics, Western Michigan University, Kalamazoo, MI 49008}

\author{V. Sih}
\affiliation{Applied Physics Program, University of Michigan, Ann Arbor, MI 48109}
\affiliation{Department of Physics, University of Michigan, Ann Arbor, MI 48109}

\date{February 03, 2015}%

\begin{abstract}
Modern electronic devices utilize charge to transmit and store information.  This leaves the information susceptible to external influences, such as radiation, that can introduce short timescale charge fluctuations and, long term, degrade electronic properties.  Encoding information as spin polarizations offers an attractive alternative to electronic logic that should be robust to randomly polarized transient radiation effects.  As a preliminary step towards radiation-resistant spintronic devices, we measure the spin properties of n-GaAs as a function of radiation fluence using time-resolved Kerr rotation and photoluminescence spectroscopy.  Our results show a modest to negligible change in the long-term electron spin properties up to a fluence of 1x10$^{14}$ (5 MeV protons)/cm$^2$, even as the luminescence decreases by two orders of magnitude.
\end{abstract}


\maketitle

The vast majority of modern technology relies on controlling electronic charge within a circuit.  Timing, location, and quantity of the charge are the fundamental parameters for logic operations.  Anything that can disrupt control of these parameters is an information processing hazard.  Radiation filled environments are a difficult challenge for electronic logic as particle collisions can randomly introduce large quantities of charge in the short term and degrade circuit properties in the long term.  Spin based logic has been proposed as an alternative that would offer novel functionality\cite{Zutic2004a,Huang2007e,Awschalom2009,Behin-Aein2010,Dery2011b,Zutic2011,Lee2012,Vogt2014,Chen2014} and the transmitted information should be inherently robust to short term charge effects.  A radiation-resistant spintronic device should account for bursts of induced electrical current so that it would not be damaged, while accurately measuring the quantity of spin current.  An ideal device would use a pure spin current such that charge and spin behavior are completely decoupled.
	
In order to fabricate a radiation-resistant spintronic device, a material must be chosen that is relevant for spintronics applications and largely maintain its spin dependent properties after irradiation.  In this paper, we explore the effects of irradiation on the spin properties of bulk Si-doped n-GaAs samples cleaved from an off-the-shelf wafer.  We expose several samples to proton irradiation, and then characterize them using photoluminescence (PL) and gamma spectroscopy.  We then perform resonant spin amplification (RSA) using pump-probe Kerr rotation to extract the spin dependent parameters.  Our results show that the spin lifetime and g-factor of bulk n-GaAs, doped near the metal-to-insulator transition, is largely unaffected by proton irradiation.  We recommend n-GaAs for further study as a candidate for radiation-resistant spintronic devices.
	
All samples were cleaved into 4 mm x 4 mm x 0.5 mm chips from the same bulk Si-doped n-type GaAs wafer.  The parent wafer was 2 in diameter x 0.5 mm thick, with the following manufacturer specifications: carrier concentration of (4.3-6.2)x10$^{16}$ cm$^{-3}$, mobility of (3450-3880) cm$^2$/V$\cdot$s, and resistivity of (2.8-3.9)x10$^{-2}$ Ohm$\cdot$cm.  One sample was set aside as a reference.  Six samples were irradiated at Western Michigan University's 6.0 MV Van de Graaf accelerator facility with a selection of 5.0 MeV H+ ion fluences:  2.5x10$^{12}$, 1x10$^{13}$, 1x10$^{14}$, 1x10$^{15}$, 1x10$^{16}$, and 1x10$^{17}$ protons/cm$^2$.  For comparison, equipment in a satellite monitoring ocean features operating at a 1334 km orbit above the earth, with a 63 deg angle of inclination and 100 mil Aluminum shielding, will experience a fluence of approximately $10^{11}$ protons/cm$^2$ in one year with energies distributed from 0.1 MeV to 1 GeV, peaked around 10 MeV.\cite{Johnston2010}  It should be noted that different proton energies lead to different forms of damage.  Low energy protons (~1 MeV) can become lodged within a material inducing swelling near a surface,\cite{Schindler1989} while high energy protons can induce nuclear reactions.\cite{Zakharenkov1990}

\begin{figure}[t]
\includegraphics[scale=1]{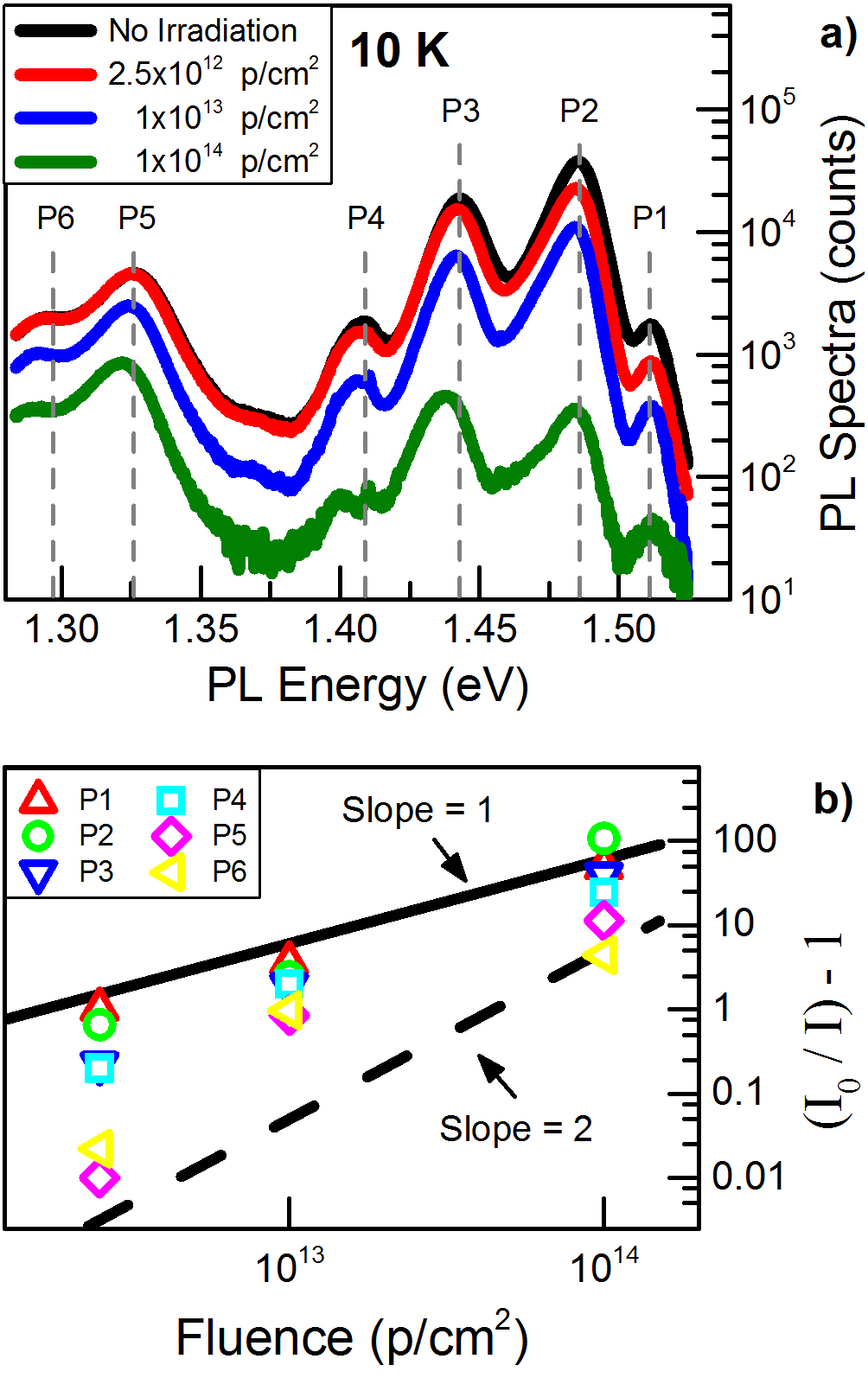}
\caption{a) Photoluminescence (PL) measurements of irradiated and reference samples.  b) Fluence dependent trends of PL maxima labeled P1-P6 in Fig. 1a, plotted using the function $\left(I_0/I \right) - 1$. $I_0$ is the reference, non-irradiated PL intensity.  $I$ is the intensity at a given fluence.  Black lines are guides to the eye.}
\label{f1}
\end{figure}

We performed PL measurements on all samples using the 1.96 eV emission of a HeNe laser at 1 W/cm$^2$ intensity for excitation.  All samples were mounted in a liquid-He cooled cryostat with data collection at 10 K.  A liquid nitrogen cooled charge-coupled device array (CCD) and grating spectrometer were used to detect the PL.  Figure \ref{f1} shows PL collected from samples up to the 1x10$^{14}$ protons/cm$^2$ fluence.  We observed negligible PL signal from samples exposed to higher fluence.

\begin{figure*}[t]
\centering
\includegraphics[width=\textwidth]{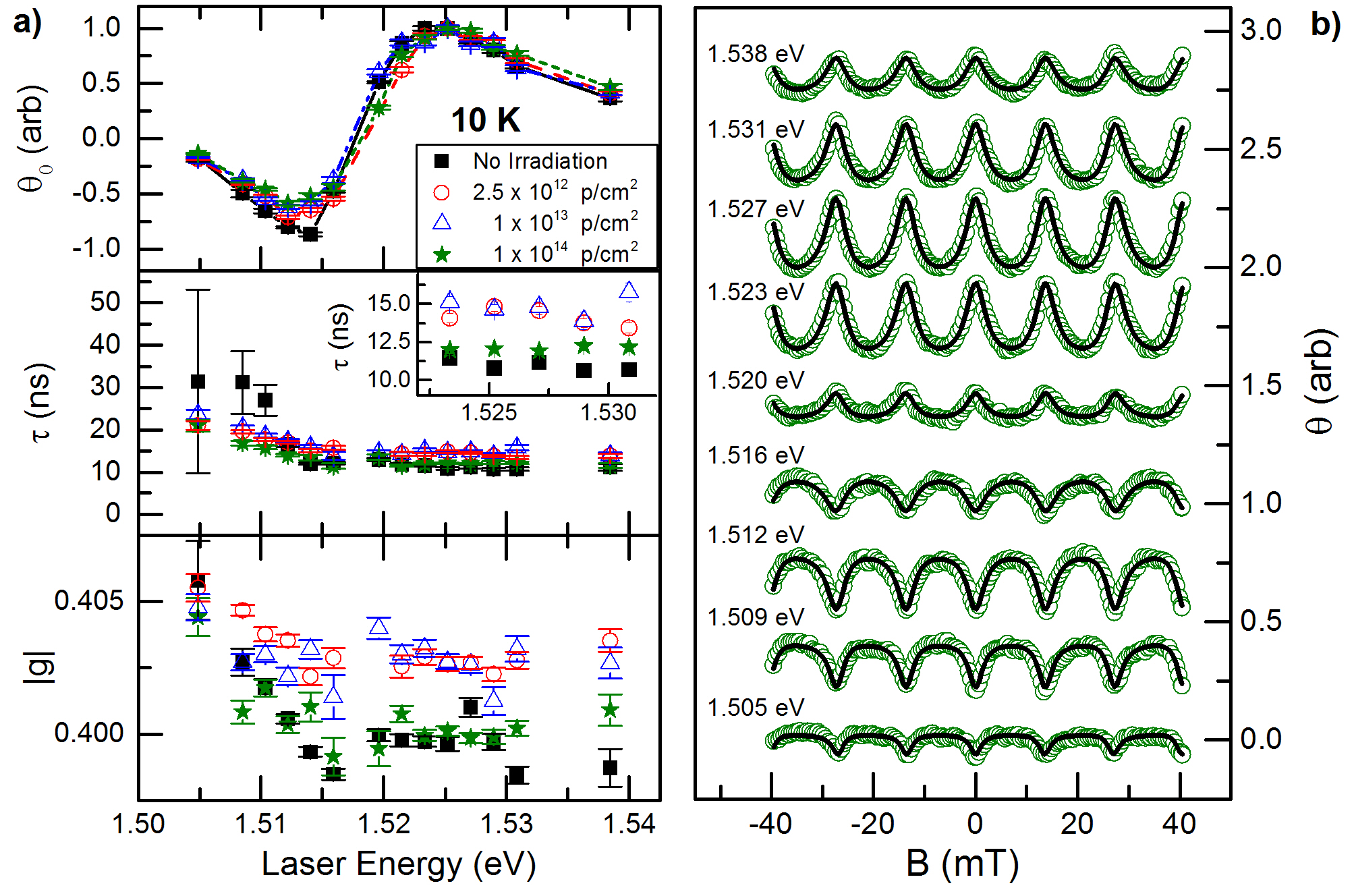}
\caption{a) Fitting parameters as a function of laser energy and fluence, extracted from Eq. \ref{fitFunc}.  $\theta_0$ values are normalized and lines are guides to the eye.  The inset is a zoom of the $\tau$ dependence on laser energy for clarification.  b) Data from the 1x10$^{14}$ protons/cm$^2$ fluence sample as a function of applied magnetic field (green circles) with fits (black lines).  Each curve is labeled with the corresponding laser energy on the left-hand side of the figure.}
\label{f3}
\end{figure*}

In Fig. \ref{f1}a, we label the observed PL peaks P1 through P6 and plot the relative intensity of each peak and their dependence on fluence in Fig. \ref{f1}b.  PL maxima P1, centered at 1.512 eV, is attributed to the band-to-band transition.  This is supported by the spin dependent data shown in Fig. \ref{f3}a where the strongest negative polarization and general change in spin dependent behavior occurs at $\sim$1.512 eV.  P2, centered at 1.485 eV, is attributed to excitonic and shallow acceptor transitions. Neither P1 nor P2 shift their center position with increasing fluence.  PL maxima at P3, P4, P5, and P6, centered at 1.443 eV, 1.408 eV, 1.326 eV, and 1.297 eV respectively before irradiation, shift to lower energy with increasing fluence.  P3 and P4 are potentially phonon replicas of P1 and P2 while P5 and P6 are likely dominated by impurities.  P4 also appears to split into two separate peaks with increasing fluence, possibly from the introduction of new defect states or the unequal degradation of multiple PL transitions.  For P4, we only analyzed the lower energy peak.

Figure \ref{f1}b follows the degradation of the PL maxima with fluence, comparing the intensity before and after irradiation, using the equation $\left[ \left( I_0/I \right) - 1 \right] = K \phi^m$, where $I_0$ is the intensity prior to irradiation, $I$ the intensity after, $\phi$ the corresponding fluence, and $K$ the degradation coefficient.\cite{Khanna1996}  The exponent $m$ is determined by the type of damage introduced to the sample.  The introduction of mid-gap states leads to a linear dependence on fluence, while a high concentration of radiation-induced complex formations leads to a quadratic dependence.  Figure \ref{f1}b shows that the behavior for all peaks ranges from linear to modestly super-linear, as expected for 5 MeV proton irradiation in n-type GaAs.\cite{Khanna1996}

\begin{figure}[t]
\includegraphics[scale=1]{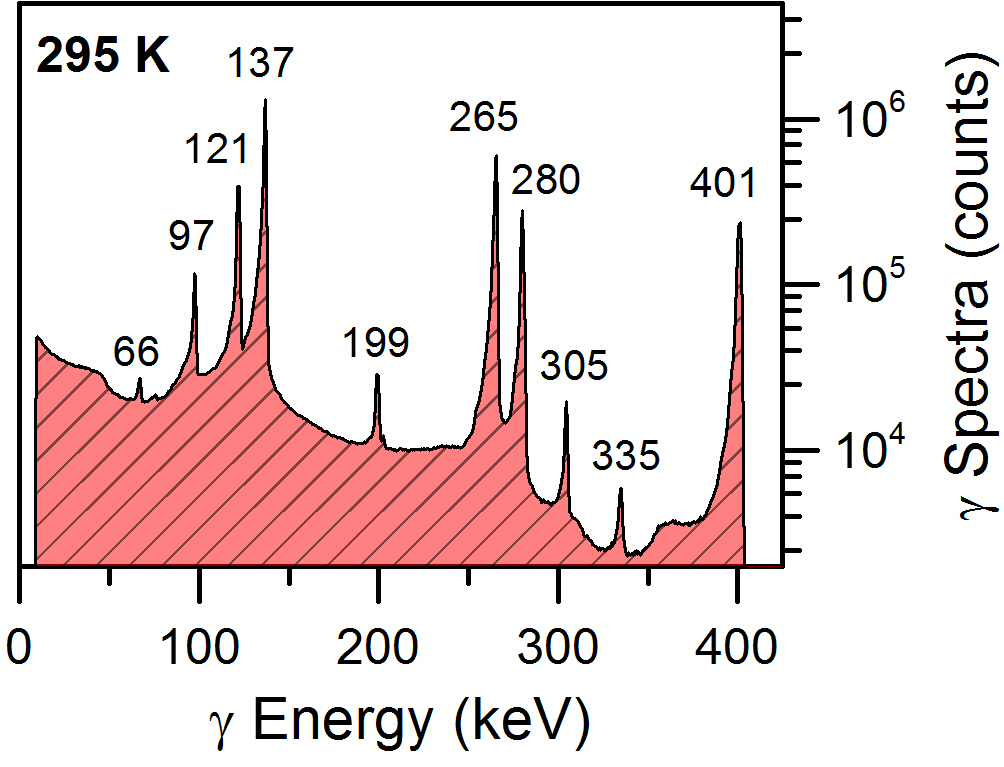}
\caption{Gamma spectra taken from the 1x10$^{17}$ protons/cm$^{2}$ fluence sample.  Spectral peaks are identified with their corresponding energy in keV.}
\label{f2}
\end{figure}

After proton irradiation, the samples exhibited increased radioactivity, which persisted in the sample exposed to the highest fluence, 1x10$^{17}$ p/cm$^2$, even after several weeks.  In order to determine the source of radioactivity, we collected gamma spectra on that sample, at room temperature, at the University of Michigan Advanced Physics Teaching Laboratory using a high purity Ge solid state detector.  The data, shown in Fig. \ref{f2}, reveals the likely culprit to be decay of $^{75}$Se.  The only stable isotope of Arsenic, $^{75}$As, has 100\% natural abundance.  There exists a possible nuclear reaction between a proton and an $^{75}$As neutron, with an energy barrier of only 1.6 MeV and reaction product $^{75}$Se.\cite{Zakharenkov1990}  The half-life of $^{75}$Se is 120 days and decays back to $^{75}$As.  Our gamma spectra match what was reported, for the same energy range, in Ref. 8\nocite{Zakharenkov1990} and measured spectral peak energies were compared to the NuDat 2.6 database\cite{Aleja2014} for decay radiation.  The vast majority of the signal can be accounted for by $^{75}$Se decay to $^{75}$As.  

We performed Kerr rotation measurements utilizing a tunable Ti:Sapphire pulsed laser.  Each laser pulse was 3 ps full-width at half-max (FWHM) in time and generated at a 76 MHz repetition rate.  A beam splitter divided each pulse onto two paths:  1) a pump path with mechanical delay line and 50 kHz photo-elastic modulator (PEM); and 2) a linearly polarized probe path with an optical chopper operating at 500 Hz.  Upon reflection from the sample surface, the orientation of the probe's linear polarization was detected by a Wollaston prism, photo-diode bridge, and two cascaded lock-in amplifiers.  The first lock-in was synchronized to the PEM and the second was synchronized to the optical chopper.  The mechanical translation stage with a retroreflector delayed the probe pulse relative to the pump pulse allowing for picosecond time resolution over a 13 ns scannable range.  For all spin-dependent measurements reported here, a 0.4 mW probe was delayed by 12.96 ns relative to a 3.0 mW pump.  Both beams had 30 $\mu$m FWHM spot diameters and overlapped at the sample surface.  All samples were mounted in a liquid-He cooled cryostat with data collection at 10 K.

Spin ensembles were generated by exploiting the optical selection rules of GaAs.\cite{Zutic2004a}  The PEM modulated the pump between right- and left- circular optical polarization to photo-generate the spin polarization of carrier ensembles along, or against, the optical path.  We performed resonant spin amplification (RSA) in the Kerr rotation geometry on samples up to the 1x10$^{14}$ p/cm$^2$ fluence, extracting the dephasing time and g-factor.\cite{Kikkawa1998}  

RSA is the byproduct of cumulatively measuring several non-interacting spin ensembles, each generated by their own pump pulse, separated in time by the laser repetition period.  Each spin ensemble undergoes Larmor precession from an externally applied magnetic field and dephasing from various spin scattering mechanisms.  If the ensemble dephasing time is comparable to, or longer than, the laser repetition period, multiple ensembles will be detected by the probe pulse.  If the precession frequency is an integer multiple of the laser repetition frequency, the collective measurement of the spin ensembles will yield a constructive interference maxima.  By fixing the delay time and tuning the applied magnetic field, several instances of constructive maxima can be observed.  

The derivation of a quantitative model for RSA is rather straightforward:  utilize a model for single ensemble spin precession and dephasing, then sum over a large number of ensembles generated at the laser repetition rate.  The governing equation used for a single spin ensemble is
\begin{equation}
\frac{\partial \mathbf{S}(t)}{\partial t} - \mathbf{\Omega}\times \mathbf{S}(t) + \frac{\mathbf{S}(t)}{\tau} = \mathbf{S_0} \delta(t)
\label{govEq}
\end{equation}
where $\mathbf{S}(t)$ is the spin polarization per unit volume, $\mathbf{S_0}$ is the polarization at time 0, $\tau$ is the dephasing time and $\delta(t)$ is the Dirac-delta function.  $\mathbf{\Omega}$ is the Larmor precession frequency defined as $\mathbf{\Omega} = \mu_B g \mathbf{B} / \hbar$ where $\mu_B$ is the Bohr magneton, $\hbar$ is the reduced Planck's constant, $g$ is the Lande g-factor, and $B$ is the applied magnetic field.  For our experimental setup, we define $\mathbf{B}=B\hat{x}$  and $\mathbf{S_0}= S_0 \hat{z}$, where $\hat{z}$ is the optical axis.  We connect the spin polarization to the amount of Kerr rotation using the relation $\theta \propto S_z$, where $\theta$ is the angle between the initial and final linear probe polarizations and $S_z$ is the component of the spin polarization along the optical axis.\cite{Giri2012}  The resulting equation for the time dependent Kerr rotation is
\begin{equation}
\theta(t,\Omega) = \theta_0 H(t) e^{-t/\tau} cos\left(\Omega t \right)
\label{simpleSolution}
\end{equation}
where $H(t)$ is the Heaviside step function.  We take Eq. \ref{simpleSolution} and then sum over an infinite number of pump pulses generated at the laser repetition rate,  $t_R$, to arrive at Eq. \ref{seriesSolution}.
\begin{equation}
\theta(t,\Omega) = \theta_0 \sum_{n=0}^{\infty}  H(t_n) e^{-t_n / \tau} cos\left[ \Omega \, t_n \right]
\label{seriesSolution}
\end{equation}
where $t_n = t + n \, t_R$.  By assuming that $t>0$, the sum can be evaluated as a complex geometric series yielding Eq. \ref{fitFunc} (an alternate form of Eq. 10 in Ref. 12\nocite{Glazov2008}) which was used to fit all RSA data.  
\begin{equation}
\theta(t,\Omega) = \frac{\theta_0 e^{- t / \tau} \left( cos\left[\Omega t \right] - e^{-t_R/\tau}cos\left[\Omega \left( t- t_R \right) \right] \right)}{sin^2\left[ \Omega \, t_R \right] + \left( cos\left[ \Omega \, t_R \right] - e^{-t_R/\tau} \right)^2 }
\label{fitFunc}
\end{equation}
If $\tau \ll t_R$, then Eq. \ref{fitFunc} reduces to Eq. \ref{simpleSolution} as expected.

Figure \ref{f3} shows selected RSA data with fitting curves, along with the laser energy dependence of all fit values.  Each data point was collected at least four times and then averaged.  Our fits assume a single spin species which, for our samples, should be electrons as the holes are expected to have sub-picosecond spin lifetimes\cite{Zutic2004a} and our pump modulation minimizes dynamic nuclear polarization.\cite{Trowbridge2014}  There is a distinct, and expected,\cite{Norman2014b} dependence on laser energy for the maximum rotation value $\theta_0$, stemming from the difference in indexes of circular refraction generated by the injected spin polarization.  There is also an increase in the magnitude of the dephasing time $\tau$ with decreasing laser energy below the bandedge.   Reference 14\nocite{Crooker2009} showed that such behavior could be attributed to changes in absorbed probe power, independent of wavelength,  whereby probe-induced photo-excited holes reduce the spin lifetime via the Bir-Aranov-Pikus (BAP) mechanism.\cite{Bir1975}

Unexpectedly, the measured spin properties exhibit negligible change with H+ fluence.  The magnitudes of $g$, and $\tau$, appear to increase for the lowest two fluences by less than 2\% and 40\% respectively, relative to the reference sample.  $\mid g\mid$ and $\tau$ then return to their reference values at the 1x10$^{14}$ p/cm$^2$ fluence.  The normalized amplitude of $\theta_0$ shows no clear trend and the raw values were all within a factor of 2 of each other.  This shows that the ability to optically pump and measure the circular birefringence is robust to the effects of irradiation for our experiment conditions.  The modest behavior of the spin properties is in contrast to the orders of magnitude monotonic decrease in measured PL intensity (see Fig. \ref{f1}).  For comparable fluences of 5 MeV protons to those measured here, electrical resistivity is reported to increase by several orders of magnitude.\cite{Brudnyi1980,Brudnyi1980a}  It is possible that the measured spin behavior stems from inconsistencies in the dopant level of the n-GaAs wafer, though more likely irradiation played some role as we discuss below.

Figure \ref{f3}a shows that three data points for $\tau$ of the reference sample deviate from the rest of the data set.  At the three lowest laser energies, the reference sample's spin lifetime appears to be significantly longer than the irradiated sample's with substantially higher error bars.  As all the data shown was taken with identical settings, including magnetic field resolution, subsequent runs were conducted to determine the validity of those data points.  Doubling the magnetic field resolution did shrink the error bars on the points in question, but it did not appreciably shift the central value or other fitting parameters (data not shown).  A plausible irradiation mechanism for the change of low energy lifetimes is the placement of defects within the bandgap.  Such defects can form charge traps, thereby changing the sample's carrier density and, in some cases, modestly enhancing electrical properties as has been reported for low fluences of neutron irradiation.\cite{Jorio1993}  Some of these charge traps could be introduced near the valence band, leading to an enhancement of the BAP spin dephasing.  This would explain the abrupt reduction of spin lifetime in the irradiated samples, compared to the reference sample, below the bandedge.

In conclusion, we report the robustness of (4.3-6.2)x10$^{16}$ cm$^{-3}$ Si-doped n-GaAs spin properties to 5 MeV H+ irradiation.  We confirmed sample damage from irradiation using PL and gamma spectroscopy that are analogous to data found in the literature.  There is an abrupt, though modest, drop in the spin lifetime below the bandgap, relative to the reference sample data, which may be accounted for by radiation induced charge traps near the valence band enhancing the BAP dephasing mechanism.  Our results show that GaAs, already a technologically valuable material, has promising attributes for the development of radiation-resistant spintronic devices.

This work was supported by the Defense Threat Reduction Agency Basic Research Award No. HDTRA1-13-1-0013, the National Science Foundation Materials Research Science and Engineering Center program DMR-1120923 and under Grant No. ECCS-0844908, the Office of Naval Research, and the Air Force Office of Scientific Research.  B.P. was supported in part by the Graduate Student Research Fellowship under Grant No. DGE 1256260.  

\bibliographystyle{aipnum4-1}
\bibliography{References}

\end{document}